# A glimpse of a Luttinger liquid

IGOR A. ZALIZNYAK

*Physics Department, Brookhaven National Laboratory, Upton, NY 11973, USA*

*email: zaliznyak@bnl.gov*

**The concept of a Luttinger liquid has recently been established as a fundamental paradigm vital to our understanding of the properties of one-dimensional quantum systems, leading to a number of theoretical breakthroughs. Now theoretical predictions have been put to test by the comprehensive experimental study.**

Our everyday life experience is that of living in a three-dimensional world. Phenomena in a world with only one spatial dimension may appear an esoteric subject, and for a long time it was perceived as such. This has now changed as our knowledge of matter's inner atomic structure has evolved. It appears that in many real-life materials a chain-like pattern of overlapping atomic orbitals leaves electrons belonging on these orbitals with only one dimension where they can freely travel. The same is obviously true for polymers and long biological macro-molecules such as proteins or RNA. Finally, with the nano-patterned microchips and nano-wires heading into consumer electronics, the question "how do electrons behave in one dimension?" is no longer a theoretical playground but something that a curious mind might ask when thinking of how his or her computer works. One-dimensional problems being mathematically simpler, a number of exact solutions describing "model" one-dimensional systems were known to theorists for years. Outstanding progress, however, has only recently been achieved with the advent of conformal field theory[1,2]. It unifies the bits and pieces of existing knowledge like the pieces of a jigsaw puzzle and predicts remarkable universal critical behaviour for one-dimensional systems[3,4]. In the recent study published in Nature materials[5], B. Lake *et al.* confront the theory with a comprehensive experimental study of the dynamical properties of a generic one-dimensional system – a chain of quantum spins ½ – and find spectacular agreement between their data and the theoretical predictions.

While we can't experience living in low-dimensional worlds, they are full of surprises that we can readily appreciate. One of the first things that come to mind is the spectacular argument (given by Stephen Hawking in one of his popular lectures[6]) on why life as we know it is not possible in two dimensions: The digestive tract would divide a two-dimensional animal into two separate pieces! Geometry is even more restrictive in one dimension. Here two objects can not move past one another unless they can penetrate each other; the one on the right will always remain on the right, and the one on the left will always be on the left. Hence, in addition to political implications, a clear distinction between the two fundamental types of particles, those obeying Bose and Fermi statistics, disappears in the one-dimensional world. Indeed, the difference between bosons and fermions comes into play in quantum mechanics when two particles swap places. This has no effect for the system of bosons but changes the sign of the wave function for fermions. If particles never swap places, the system's descriptions in terms of Bose and Fermi elementary excitations are equally legitimate, the choice being just a matter of convenience as the non-interacting fermions are equivalent to strongly interacting bosons and vice versa[7,8]. Capitalizing on this observation and employing the powerful machinery of quantum field theory theorists developed a new technique known as *bosonisation*[4,9] which provides a unified description of the one-dimensional world.

Discovery of conformal invariance has further revolutionized our understanding, providing a potentially complete description of all possible critical phases in 1+1 (space + time) dimensions. Perhaps, the most important is the so-called "Tomonaga-Luttinger" or, simply "Luttinger" liquid that was originally devised as a replacement for Landau

Fermi-liquid theory for one-dimensional metals[10]. Fermi-liquid accurately predicts the properties of "usual", three-dimensional metals, but fails dramatically in one dimension. A Luttinger liquid description applies to a wide class of one-dimensional quantum systems with "photon-like" elementary excitations (i.e. excitations that can carry arbitrarily small energy). The ground state of such systems is *quantum-critical*, that is, at zero temperature they are in a critical state where the fluctuations in the system are correlated on all length scales but no long-range order exists.

A Luttinger liquid theory predicts universal properties for the great variety of one-dimensional systems, including the electronic states of carbon nanotubes and nano-wires, conducting properties of conjugated polymers and fluid behaviour of Bose liquids confined within one dimensional nano-capillaries. The simplest and best studied example of the Luttinger liquid is a chain of quantum spins ½ where the energy depends on the misalignment of the nearest neighbours. Notwithstanding the fact that an exact *ansatz* solution of the problem was found by Bethe back in 1932, there was little progress in computing the physical properties that are measured in experiment, such as the dynamical spin susceptibility. Conformal field theory makes such calculations a simple exercise. The caveat, however, is that the field-theoretical predictions are only valid at sufficiently low energies and for sufficiently low temperatures[4,11]. What does *sufficiently* mean in practice? B. Lake et al. answer this question by comparing the results of their comprehensive experimental study of spin dynamics in the model spin-chain material $KCuF_3$ with the field theory predictions.

$KCuF_3$ is a fascinating material. An orbital ordering in its incipient cubic structure is such that copper *d*-orbitals carrying spin-½ overlap along *c*-direction, forming chains, while the orbitals in the neighbour chains are perpendicular and have minimum overlap. Hence, the interactions are mainly constrained to one dimension. $KCuF_3$ is an insulator; the energy required to move the electron's charge from site to site is more than an electron-volt. The only low-energy degree of freedom is the direction of the electron's spin. The virtual quantum tunnelling of the electrons between the neighbour sites favours the opposite direction of the neighbour spins. Hence, for energies below about 0.1 eV $KCuF_3$ is effectively a model system of antiferromagnetic spin-½ chains.

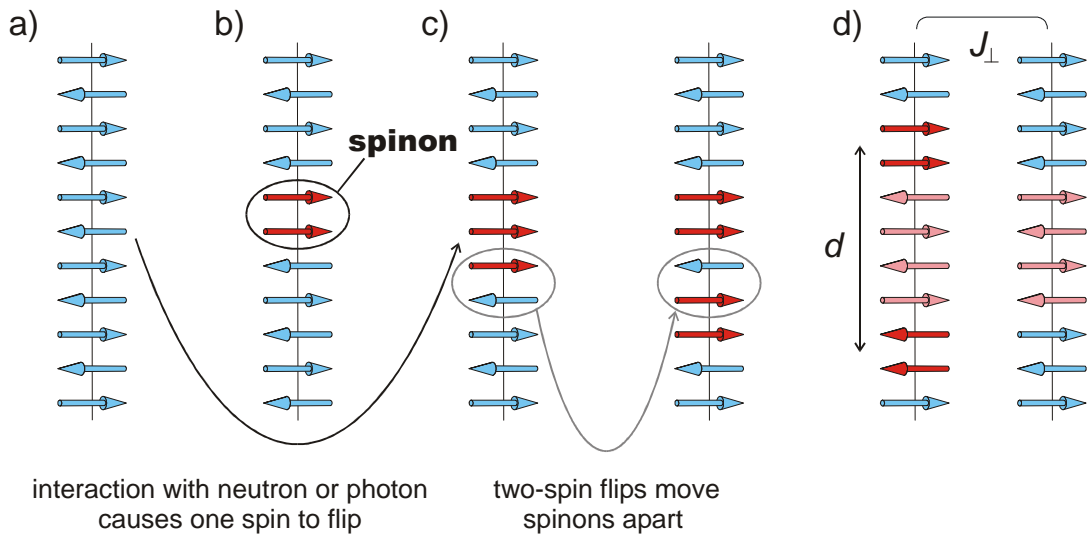

**Figure 1**. Excited modes of a 1-D quantum antiferromagnet. **a**, The ground state of a 1-D antiferromagnet with $\Delta > 1$ is simply a chain of ½ spins in which each is aligned antiparallel to its neighbours. **b**, The fundamental excitation in the antiferromagnetic spin-½ chain is a mobile domain wall that consists of two parallel spins — collectively known as a spinon (shown in red) — separating two antiferromagnetic domains. The total spin in the isolated chain being conserved only simultaneous flips of two opposite spins are permitted. Flipping the pair including one of the spins (red arrow) in a spinon by 180° causes the spinon to move by two lattice periods. Similarly, flipping a spin or a spin-pair at the end of a chain can induce the creation of a spinon, which can then propagate and carry a ½-unit of spin down the chain. **c**, Through interaction with the magnetic field of a photon or neutron from a probe beam, a single spin in the middle of a chain can be flipped. Quantum selection rules dictate that the total spin of the chain

can only change by 1, which therefore necessitates the generation of two spinons. In a single chain the spinons are free and can subsequently move apart by consecutive two-spin flips (see also left side of the panel d). **d**, In KCuF$_3$ the inter-chain coupling $J_\perp$ acts to co-align the nearest spins in the neighbouring chains. In the presence of three-dimensional ordering in the system of coupled chains, the spins between the two spinons are misaligned with equivalent sites on the neighbouring chains. Consequently, the total energy of the system increases in proportion with spinon separation, *d*, which generates a linear attractive potential between spinons. The lowest-energy two-spinon state is their bound state where spinons are confined, forming a magnon, such as the case shown on the left in the panel **c**.

The fundamental carriers of energy quanta in the spin-½ chain are the elementary excitations called spinons, Figure 1. Spinons are *fractional*: they carry spin ½ and in the physical processes involving interaction with the external electromagnetic field are always created in pairs. Spinons are *topological*: a single spinon can only appear at the topological singularity such as the chain's end, Figure 1. Spinon pairs dominate the low-energy excited states of the spin-½ chain, forming a continuum band of allowed energies whose boundaries are determined by the spinon dispersion law. While glimpses of spinon continuum were previously seen in different materials[12,13,14], the measurements of B. Lake et al. provide the most complete experimental picture of this continuum and its evolution with temperature.

Luttinger liquid phase being critical, even a tiny residual interaction between the chains in real material leads to the three-dimensionally ordered ground state where the spinons are bound in pairs forming the coherently propagating spin-1 magnons, Figure 1(d). Only in the absence of the inter-chain interaction, i.e. for a single chain, the Luttinger liquid is the true ground state. For the generalized XXZ chain where the spin-spin coupling is anisotropic so that the energy costs of misaligning the spins z-components and their x- and y-components are different, the Luttinger liquid phase occupies the whole interval of the ground states between the Ising ferromagnet and the Ising antiferromagnet, Figure 2. Capitalizing on the exact solution for the single chain the quantum field theory approach has been successful in describing the system of the weakly coupled chains whose ground state is ordered[4,9,11] as in $KCuF_3$, Figure 2. It predicts that at higher energies *or* at temperatures above the three-dimensional ordering the spin dynamics is still governed by the Luttinger liquid theory, while at low energies *and* temperatures it is similar to that of a classical three-dimensional antiferromagnet. In the intermediate, cross-over energy region, an appearance of the unusual longitudinal mode corresponding to the length oscillation of the antiferromagnetic order parameter is predicted. By carefully studying the spin-excitation spectrum in $KCuF_3$ in the ordered phase B. Lake et al. put all of these predictions to test.

It is difficult to overemphasize the importance of the spectacular measurements reported by B. Lake et al. This study provides experimental credibility to the beautiful abstract mathematics of the conformal field theory and establishes the limits of its applicability in real systems. It also furnishes a spectacular demonstration of how the quantum-critical Luttinger liquid state governs the dynamical properties of the systems within the whole region of the phase diagram (cf. Figure 2). It is a perfect experimental illustration for a textbook on applications of quantum field theory in condensed matter physics[4].

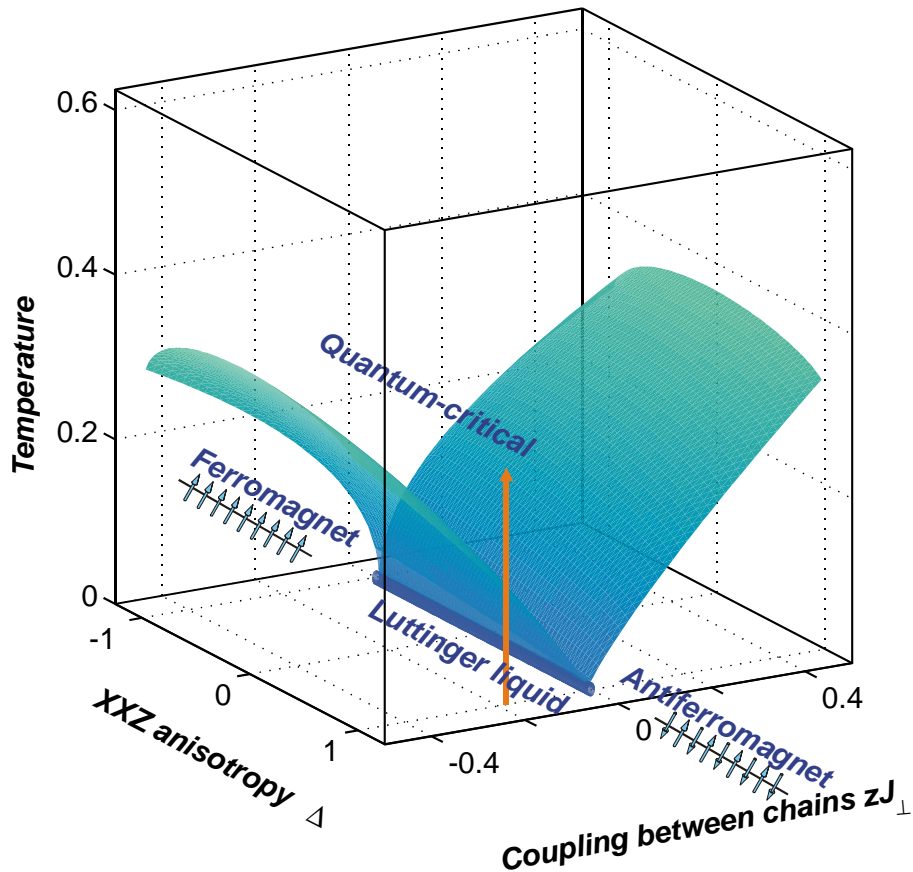

**Figure 2.** Phase diagram for the system of weakly coupled XXZ spin ½ chains. In XXZ chain the relative importance of the misalignment of the spins z-components and that of their x and y components are different; this is quantified by the anisotropy parameter $\Delta$. The surface shows temperature of the three-dimensional antiferromagnetic ordering[11] resulting from the weak coupling between the chains as a function of the strength of this coupling, $zJ_\perp$, ($z$ is the number of the neighbour chains, $z = 4$ for $KCuF_3$) and the uniaxial anisotropy of the spin-spin interaction in the chain, $\Delta$. The Luttinger liquid phase occurs at $T = 0$ in the region $-1 < \Delta \leq 1$ and for $zJ_\perp = 0$. For $\Delta < -1$ the system is an Ising ferromagnet, for $\Delta > 1$ it is an Ising antiferromagnet. The red arrow shows the line studied by Lake *et. al.* in $KCuF_3$, which is a Heisenberg antiferromagnet with $\Delta = 1$.

**References**

1. Belavin, A. A., Polyakov, A. M., and Zamolodcchikov, A. B., Infinite conformal symmetry in two-dimensional quantum field theory, Nucl. Phys. B 241, 333 (1983).

2. Cardy, J. L., Conformal invariance and universality in finite-size scaling, J. Phys. A: Math. Gen. 17, L385 (1984).

3. Affleck, I., Exact critical exponents for quantum spin chains, non-linear σ-models at $\theta = \pi$ and the quantum hall effect, Nucl. Phys. B 265, 409 (1986).

4. Tsvelik, A. M., Quantum Field Theory in Condensed Matter Physics, Second edition (Cambridge University Press, Cambridge, 2003).

5. Lake, B., Tennant, D. A., Frost, C. D. & Nagler, S. E. *Nature Mater.* **4,** 329–334 (2005).

6. C. L. Broholm, private communication.

7. Mattis, D. C., and Lieb, E. H., Exact Solution of a Many-Fermion System and Its Associated Boson Field, Journ. Math. Phys. 6, 304 (1965); Mattis, D. C., New wave-operator identity applied to the study of persistent currents in 1D, *ibid.* 15, 609 (1974).

8. Regnault, L.-P., Zaliznyak, I. A., and Meshkov, S. V., Thermodynamic properties of the Haldane spin chain: statistical model for the elementary excitations, J. Phys.: Condens. Matter 5, L677 (1993).

9. Gogolin, A. O., Nersesyan, A. A., Tsvelik, A. M., Bosonisation and Strongly Correlated Systems, (Cambridge University Press, Cambridge, 1998).

10. Haldane, F. D. M., "Luttinger liquid theory" of one-dimensional quantum fluids, J. Phys. C: Solid State Phys. 14, 2585 (1981).

11. Bocquet, M., Essler, F. H. L., Tsvelik, A. M., and Gogolin, A. O., Finite-temperature dynamical susceptibility of quasi-one-dimensional frustrated spin-½ Heisenberg antiferromagnets, Phys. Rev. B. 64, 094425 (2001).

12. Zaliznyak, I., Woo, H., Perring, T. G., Broholm, C. L., Frost, C. D., and Takagi, H., Spinons in the strongly correlated copper oxide chains in SrCuO2, Phys. Rev. Lett. 93, 087202 (2004).

13. Stone, M. B., Reich, D. H., Broholm, C., Lefmann, K., Rischel, C., Landee, C. P., and Turnbull, M. M., Extended Quantum Critical Phase in a Magnetized Spin-½ Antiferromagnetic Chain, Phys. Rev. Lett. 91, 037205 (2003).

14. Zheludev, A., Raymond, S., Regnault, L.-P., Essler, F. H. L., Kakurai, K., Masuda, T., and Uchinokura, K., Polarization dependence of spin excitations in BsCu2Si2O7, Phys. Rev. B 67, 134406 (2003).

**Acknowledgements**

I gratefully acknowledge numerous illuminating discussions with A. Tsvelik, R. Konik, F. Essler, A. Zheludev and L. Passell. The work at BNL was supported by the Office of Science, U.S. Department of Energy.